\documentclass[conference,compsoc]{IEEEtran}
\usepackage{graphicx}
\ifCLASSOPTIONcompsoc
  \usepackage[nocompress]{cite}
\else
  \usepackage{cite}
\fi
\ifCLASSINFOpdf
\else
\fi
\begin{document}
%
\title{System Design Considerations For Internet Of Things (IoT) With Category-M Devices In LTE Networks}

\author{\IEEEauthorblockN{Gurudutt Hosangadi}
\IEEEauthorblockA{Nokia\\
600-700 Mountain Ave,Murray Hill \\
NJ 07974-0636, USA \\
Email:gurudutt.hosangadi@nokia.com}
\and
\IEEEauthorblockN{Dandan Wang}
\IEEEauthorblockA{Nokia\\
600-700 Mountain Ave, Murray Hill \\
NJ 07974-0636, USA \\
Email:dandan.wang@nokia.com}
\and
\IEEEauthorblockN{Anil Rao}
\IEEEauthorblockA{Nokia \\
12004 173rd Pl NE \\
Redmond, WA 98052, USA \\
Email:anil.rao@nokia.com}}


%


\maketitle

\begin{abstract}
    Successful network deployment of the Internet of Things (IoT) requires many critical system design considerations. This paper highlights how an LTE system supporting Cat-M devices can be engineered to deal with the numerous constraints the 3GPP standard imposes for this new device type. Fundamental changes to the control channels, control and data timing relationships, the need to support half-duplexing, and variable repetition lengths pose non-trivial challenges, particularly when attempting to satisfy the critical coverage KPI for Cat-M devices while at the same time preserving the capacity KPI for legacy LTE devices. In addition, the nature of IoT traffic is fundamentally different than legacy LTE data, requiring changes to existing system parameters and MAC algorithms.  Finally, we will touch upon the topic of supporting voice over IP traffic on Cat-M devices and the challenges therin.
\end{abstract}


%
\IEEEpeerreviewmaketitle

\section{Introduction}
The internet of things (IoT) buildup is well underway. The number of M$2$M and Narrowband-IoT devices is expected to reach $1$ billion by $2020$ \cite{statistica_research}. These devices cover a wide range of applications:  wearable devices, connected home appliances, remote sensing for utilities and smart cities to name a few. These massive number of devices communicating without human intervention constitute to what is commonly called as machine type communication (MTC) or referred to as IoT. Modern wireless cellular networks such LTE (Long Term Evolution) based on 3rd Generation Partnership Project (3GPP) are aptly placed to be an enabler of massive MTC. This is due to its all-inclusive-all-IP (internet protocol) architecture, built in security, scalable traffic management capabilities and high spectral efficiencies.
The traffic profile and requirement of IoT differs vastly from that of traditional mobile devices already supported in LTE cellular networks. Key differences include smaller traffic packet sizes and massive number of such devices. To support this change, $3$GPP standards have been enhanced with new features. See \cite{alberto,36_888} for a detailed coverage of the added features.

As we will show, the unique characteristics of IoT devices pose challenges to system design. 
In this paper, we focus on system design considerations of IoT devices and present some insights into performance based on our simulation results specifically focusing on the CaT-M feature within LTE standards. 


 

\section{Inputs to System Design and Challenges}
There are $3$ main inputs to consider while considering LTE system design support for MTC devices:
\begin{enumerate}
	\item{Requirements when MTC devices are introduced into the system}
	\item{Constraints imposed by $3$GPP}
	\item{New MTC specific features provided by $3$GPP standards}
\end{enumerate}
It is important to understand each of the above points before delving into the system design aspects as it will be evident that there are trade-offs to be considered.
\subsection{Requirements}
MTC devices have the following set of requirements: 
\begin{itemize}
	\item{The system should be able to support a massive number of MTC devices. As explained above the number of MTC is expected to jump significantly from the current levels.}
	\item{Introduction of MTC devices into the system should have minimum impact on the operation of legacy devices. Current users expect the quality of service (QoS) to be maintained for the traditional voice and data calls while having access to new MTC related services.}
\item{MTC devices have a low cost.} 
\item{MTC devices meet appropriate performance targets for each use case.  One of the key requirements is enhanced coverage \cite{alberto} compared to legacy devices. For example, a sensor in the basement of a home monitoring utility use should be able to transmit its reports to a nearby base station by overcoming the associated penetration losses. In another use case where the goal is to support VoIP services, meeting latency requirements also becomes important.} 
\end{itemize}

\subsection{Constraints}
$3$GPP imposes the following key constraints for support of MTC devices in a LTE systems:
\begin{itemize}
	\item{Half-duplexing}
	\item{Reduced operating bandwidth}
	\item{Reduced maximum transmit power}
	\item{Limit number of radio frequency chains to $1$.}
\end{itemize}
All of the above constraints were introduced for Category $0$ UEs and were extended to also apply to MTC devices with the goal of meeting the "low cost" requirement of MTC devices. There are also a few new features introduced by $3$GPP, which we will discuss later, which will allow lowering the cost further.

\subsection{Challenges}
Some of the key challenges experienced while introducing MTC devices into an LTE system are as follows:
\begin{itemize}
	\item{It can be observed that while the one of the key requirements is enhanced coverage, the constraint of reduced maximum transmit power makes achieving this more difficult. }
	\item{Although MTC devices are allocated a small fraction of the bandwidth, minimizing impact on the performance of legacy devices is another challenge given that a massive number of such MTC devices need to be supported.}
	\item{Finally if any latency sensitive services (such as VoIP) are to be supported, then half-duplexing hinders the latency objectives.}
\end{itemize}
We will discuss in more details some of above challenges in a later section.

\subsection{New features}
$3$GPP introduced in Release $13$ new features \cite{alberto} to enable MTC devices in LTE systems. The key ones are are follows:
\begin{itemize}
	\item{Narrowband operation: $3$GPP standards allows a MTC device to monitor and process a narrow bandwidth ($1.4$MHz for Cat.M$1$ and $200$KHz for Cat.NB$1$) within the available bandwidth;}
	\item{Rel $13$ introduced a mechanism of repetition (upto $256$) for MTC devices which is similar to transmission time interval (TTI) bundling in Rel $8$ (up to $4$ repeats) intended for voice over IP (VoIP) packets where consecutive TTIs are used to transmit the same packet. }
\end{itemize}

\section{System Design Guidelines}
In this section, we will discuss in more detail the different aspects of the system design to support Cat-M devices.

\subsection{Narrowband location}
\begin{figure}[htbp]
\centerline{\includegraphics[height=20mm,width=90mm]{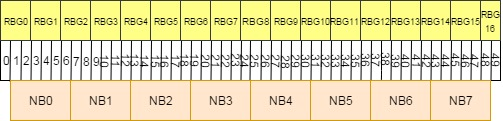}}
\caption{Narrowband reservation for CaT-M devices}
\label{fig_narrowband}
\end{figure}

The location of the narrow bandwidth used for MTC devices and the alignment of this bandwidth with resource block groups (RBGs) used by the scheduler for legacy traffic is important. For example, Figure \ref{fig_narrowband} illustrates the location of narrowbands for $10$MHz. Note that RBG0-15 has 3 PRBs while RBG 16 only has 2 PRBs. Standards requires that MTC devices are allocated $6$ consecutive PRBs which are illustrated as any one of NB$0$ through NB$7$. If we use any of NB$0$ through NB$6$, then a total of $9$ PRB cannot be used for RBG allocation. However, if we use NB$7$, fewer PRBs ($8$) cannot be used for RBG allocation.

\subsection{Repetitions and HARQ}
The hybrid automatic repeat request (HARQ) transmissions feature has been available in legacy LTE systems while repetitions were introduced in later releases for MTC devices. Both have their benefits and drawbacks. 

With HARQ, to improve the coverage, a large number of retransmission may be required. This will consume extra resource on control channel due to the grant and feedback acknowledgment. Further, the half duplexing nature of the Cat-M devices leads to large latencies and makes scheduling multiple HARQ more challenging. The benefit of HARQ is that it is adaptive i.e. there will not be additional transmissions if the packet is successfully decoded at the receiver.
Using repetitions for MTC devices uses fewer control resources and does not suffer as much from latency issues compared to using HARQ. However, repetition sizes are fixed once tranmissions for a packet begins resulting in less flexibility.

\subsection{Dormancy Timer and DRX setting impact on RACH}
 Since MTC devices are expected to carry very small amounts of data for each device (unlike legacy devices), frequent random access channel (RACH) requests will result in increased overhead especially if large repetitions are required to enhance RACH coverage. While there is no control on the number of initial RACH access requests, proper setting of dormancy time and DRX cycles can provide a good balance between subsequent RACH overhead and UE battery savings.

 The dormancy timer aims to remove the radio resource control (RRC) connection of the UE device from the evolved Node B (eNB) and thus, once it has new data to transmit or receive, it has to go through the RACH process again with the benefit of saving more UE battery. DRX, on the other hand, maintains the UE's Radio Resource Control (RRC) connection and avoids  the RACH procedure. It saves battery by turning off some portion of the radio frequency chain during the DRX off cycle and wakes up periodically to monitor the resource allocation. If there are only very small and infrequent packets sent/received by Cat-M devices, setting a smaller dormancy timer may be more desirable than configuring a DRX on/off pattern.

\subsection{Channel State Information (CSI) and Scheduling Request (SR)}
In legacy LTE systems, one mechanism for eNB to acquire downlink channel statistics of UEs is through channel quality indicator (CQI) reporting. However Cat-M devices due to the constraint of half duplexing, during the TTIs that a UE reports CQI on uplink control channel (PUCCH), this UE will not be eligible to receive any information on the downlink data and control channels. This will impact downlink throughput, especially when PUCCH has to use a large number of repetitions in a coverage limited scenario. There is a similar impact when the UE reports SR on PUCCH. Thus it may be better to configure either a relatively large CQI period to reduce downlink impact or use aperiodic CQI multiplexed with uplink data to reduce grant impact.

\subsection{HARQ process management}
\begin{figure}[htbp]
\centerline{\includegraphics[height=30mm,width=80mm]{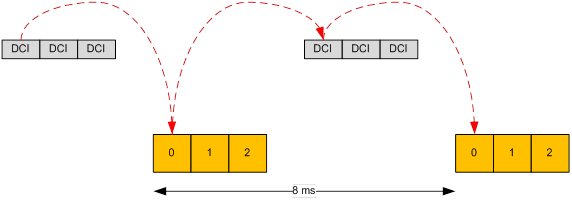}}
\caption{HARQ Timing}
\label{fig_harq_timing}
\end{figure}

Like legacy LTE devices, Cat-M devices can support multiple HARQ process at the same time independently but are subject to the constraints of half-duplexing which limits the number of HARQ processes. 
For example, Figure \ref{fig_harq_timing} shows that due to the constraints and configured repetition lengths (RLi=1), in UL a maximum of $3$ HARQ processes only may be possible if the goal is to have HARQ transmissions every $8$ms.

\subsection{Grant Channel (MPDCCH) Configuration}
In legacy LTE, the maximum aggregation level (AGL) of PDCCH is $8$ while it has been increased to $24$ for Cat-M devices. As a result, the number of grants that can be supported per TTI is smaller. Thus, there is a trade off of coverage of MPDCCH and the number of grants can be sent per TTI. For example if a high AGL (e.g. $24$) is used for a coverage limited UE, only one user can be served in a given TTI either in the uplink or downlink direction.

\subsection{Use of Closed Loop Algorithms}
As illustrated in \cite{36_888}, the traffic pattern of Cat-M is very unique in the sense that Cat-M devices wake up very infrequently and then send/receive a very small amount of data. As such, Cat-M devices are expected to wake from RRC IDLE state when new data arrives and as a result it is very difficult for any closed loop algorithms to converge and so open loop algorithms may suffice. 

\section{Results}
	In this section, we present some sample simulation results aimed at providing insight into how system design and configuration can affect performance of MTC devices.

\subsection{Assumptions and Simulation Tool}
We focus here on the use case where MTC devices are being served using the Cat.M$1$ feature support in LTE. Further it is assumed here that $1.4$MHz bandwidth is allocated and is at a fixed location. Table \ref{table_assump} lists some of key assumptions that are applicable to the results discussed.

\begin{table}
    \begin{tabular}{|l|l|}
    \hline
        {\bf Parameter} & {\bf Assumption} \\ 
    \hline
    \hline
        Frequency Bands & $2$GHz \\
        \hline
        Macro inter-site distance & $500$m \\
\hline
    Shadow model & Shadow fading std. dev = $8$ dB \\
    \hline
    Cat-M UE traffic &	$1$K bits, mean/min reading time = $10$s/$2.5$s \\
        \hline
        Dormancy timer & $2$ seconds \\
    \hline
    Fading Channel  &	ETU $3$km/hr \\
        \hline
    Macro eNB antenna &	$17$ dBi gain \\
        & Vertical pattern: $10^{\circ}$@$3$dB beamwidth \\
        & SLA = $20$ dB, downtilt =$15^{\circ}$ \\
     \hline
    Body and cable loss	& $1$ dB (data terminal) \\
    \hline
    Mobile antenna	& Omnidirection; -3 dBi gain \\
    \hline
    eNB Tx power & 	$2 \times 20$W \\
    \hline
\end{tabular}
    \caption{Assumptions For Simulation Results}

    \label{table_assump}
\end{table}

The results presented in this section were generated using a C++ system level simulator, based on $3$GPP LTE standard, which can simulate a multi-eNB layout including effects of the wireless fading channel, propagation environment and antennas. Algorithm focus is on the Medium Access Control (MAC) layer and specifically on scheduling and HARQ management. The physical layer is abstracted (to reduce simulation run time) and the simulator supports both legacy LTE devices and CaT-M devices. For the traffic profile, full buffer, VoIP and burst traffic profiles are supported. Key performance metrics such as throughputs, latencies and statistics on Signal to noise ratio (SINR), resource consumption, number of retransmission and errors rates are available for analysis from the simulator.   

\subsection{Impact of repetition length (RL)}
\begin{figure}[htbp]
\centerline{\includegraphics[height=50mm,width=80mm]{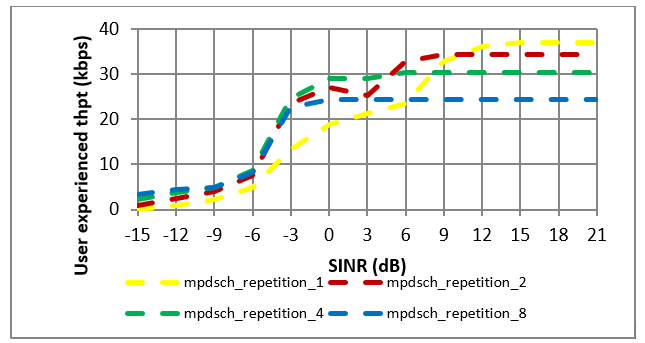}}
\caption{Impact of Repetition on User Traffic For Bursty Traffic}
\label{fig_rep_burst}
\end{figure}

We consider here the sensitivity of performance to varying the RL for a single user with burst traffic. We keep the RL of MPDCCH and PUCCH fixed at $4$ and $8$ respectively, while the RL of downlink shared channel (PDSCH) is varied. It can be observed in Figure \ref{fig_rep_burst} that increasing the RL for cell edge users definitely improves the user experienced throughput while for near cell users which have a good SINR, using a smaller repetition number makes more sense.

\subsection{Link Adaptation and Power Control}
Figures $4$A shows that the performance is insensitive to the initial block error rate (BLER) setting (e.g. $10$\%) and step size chosen (slow implies a small step size). Similarly Figure $4$B shows that closed loop power control (CLPC) provides no benefit compared to the open loop (OLPC) case. In both these cases, the average time between packets is in seconds as a result of which the MTC device transmits only $1$ measurement and then goes back to “sleep” and there is no time for any convergence.  

Figure $4$C highlights the importance of setting the correct OLPC setpoint. A lower set point allows multiple PRBs to be used which lowers the code rate and results in improved performance. Finally, Figure $4$D shows the importance of setting the initial modulation and coding scheme (MCS) on coverage. 

Note that if the traffic pattern of CaT-M devices changes (for e.g. VoIP), the system should be designed so that link adaptation and power control will improve performance. 
\begin{figure}[htbp]
\centerline{\includegraphics[height=80mm,width=80mm]{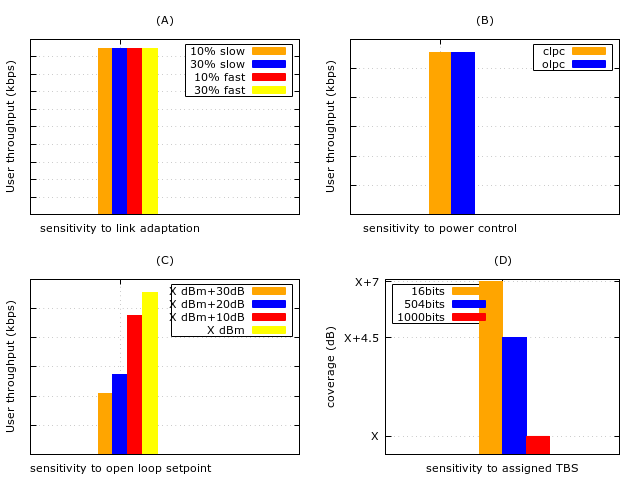}}
\caption{MTC Performance Sensitivity to Design Parameters (uplink, packet size = $1000$bits)}
\label{fig_sensitivity}
\end{figure}

\subsection{Coverage, Latency and Interference}
    Coverage is expected to be one of the key metrics for MTC devices and maximizing this will drive system design. MTC devices in coverage limited locations are expected to be transmitting at their maximum power. With this resource exhausted, the transport block sizes allocated will impact  coverage . Revisiting Figure $4$D, we can observe the sensitivity of coverage to transport block size (TBS) assignment for a packet size of  $1000$ bits.  Coverage here is defined as the maximum  path loss at which residual BLER is below $2$\%. Overall a smaller packet size provides the best coverage performance which may suffice for most MTC applications which are not expected to be time critical. 

Interference management is expected to play a key role in being able to meet the coverage requirement. One option is to reserve a narrow bandwidth solely for MTC devices in which case the interference can be low. However, if this region is used to support VoIP traffic, the interference could begin to creep up. Another option is to simply share the resources with legacy LTE devices in which case the interference levels could be quite high and meeting coverage requirements of MTC devices could become more challenging. Therefore, a system design consideration could be to consider doing some inter-cell frequency planning to ensure low interference on the reserved MTC narrowbands, which of course comes at the cost of capacity to the legacy LTE devices.

\section{VoIP support using MTC devices}
Supporting voice calls on CaT-M devices (for e.g. wearables) may be very desirable. There will now be constraints on delay on top of the coverage requirements that apply to MTC devices.
In this section, we look at different aspects of system design affecting both coverage and latency for VoIP support using CatM devices. 

\begin{figure}[htbp]
\centerline{\includegraphics[height=40mm,width=90mm]{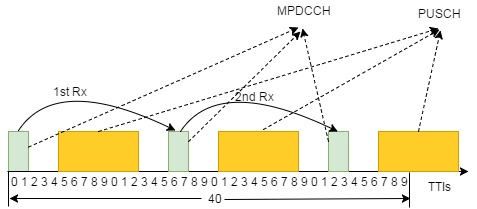}}
\caption{VoIP scheduling illustration}
\label{fig_voip}
\end{figure}

\subsection{Repetition Length, Packet Aggregation and Segmentation}
Larger RL for VoIP transmissions will increase the per link coverage but will also result in longer overall time duration for one HARQ transmission as illustrated in Figure \ref{fig_voip}. For the configured RLs, only $2$ HARQ transmissions can complete within $40$ms. As VoIP packets arrive with a fixed pattern (every 20ms at talk spurt and 160ms at silent period), in order to meet delay budgets, a larger RL implies that we may have to aggregate more VoIP packets within one HARQ transmission which requires a larger TBS. The gain achieved by the extra repetition may be offset by the lower decoding efficiency of the enlarged TBS as illustrated in Table \ref{table_mcl} which shows this effect. For a given delay budget of $200$ms, we see the supported MCL (maximum coupling loss) first increasing with higher RL but then gets worse as the RL is further increased.  Thus, there is a balance between increasing the repetition number and TBS increasing.

Segmentation is generally used in legacy VoLTE to extend the coverage. However, in the case of Cat-M, again due to the timing constraints caused by half duplexing, it is challenging to transmit multiple HARQ process at the same time especially when we use larger repetition numbers. If we segment a VoIP packets into multiple small segments, the coverage for each segment becomes better but the overall delay could be quite large.
\begin{table}
    \centering
    \begin{tabular}{|l|l|}
    \hline
        {\bf PUSCH repetition} & {\bf MCL} \\ 
        \hline
    \hline
        $8$	& $138$ \\
        \hline
        $16$	& $140$ \\
        \hline
        $32$	& $138$ \\
    \hline
\end{tabular}
    \caption{Impact Of Repetition On Coverage}
    \label{table_mcl}
\end{table}

\subsection{Impact of iBLER selection and SID packets}
In legacy LTE systems, a $10$\% iBLER target is generally used. A lower iBLER target (for example to $5$\%), will reduce the HARQ retransmission probability but will require more repetitions to support the same TBS at the same SINR. Thus, it is a trade-off between more HARQ versus repetition. As also discussed in \cite{volte}, using HARQ retransmission can achieve higher coverage than without HARQ retransmission and very large repetition. The best combination of iBLER target and RL needs further study.

In a typical voice conversation between $2$ users, during a talk spurt of one user, there will be silence insertion descriptor (SID) packets sent by the other user. Thus, this user has a voice packet to transmit at the same time it has to receive SID packets which is not allowed due to half duplexing. This makes the VoIP scheduling more challenging. Another difficulty is that we may not be able to use the same fixed TBS for VoIP packets anymore as SID packets happen less frequently compared with voice packets and whenever SID packets arrive, more aggregation of voice packets will happen.

\section{Conclusions}
We have outlined the numerous system design aspects which must be considered to successfully deploy an LTE network supporting CaT-M MTC devices. The numerous constraints as well as additional coverage/power-saving features the $3$GPP standard has included for such devices poses significant challenges in integrating support for such devices in an LTE network while minimizing the KPI impact to existing smartphone and other high performance data-centric devices. It has been shown that careful selection of system parameters such as the Cat-M dormancy timer, the number of HARQ transmissions and repetition factor involves many different tradeoffs, particularly between coverage and latency and also the capacity impact to the legacy LTE network. We have demonstrated that link adaptation features such as closed loop rate control and closed loop power control need to be revisited based on the nature of MTC traffic. It is important to highlight such considerations so that an operator can tailor the parameters and scheduler design aspects to achieve the desired trade-offs inherent in introducing MTC devices into an existing LTE network.

\end{document}